\documentclass[apjl]{emulateapj}
\usepackage{amsmath}
\usepackage{comment}
\usepackage{ifthen}
\usepackage{hyperref}
\usepackage{graphicx}
\usepackage{epstopdf}
\usepackage{epsfig}
\tabletypesize{\scriptsize}
\newboolean{emulateapj}
\setboolean{emulateapj}{true}
\newboolean{astroph}
\setboolean{astroph}{true}
\defcitealias{KeplerHandbook:2009}{\textit{The Kepler Instrument
    Handbook} (2009)}

\shortauthors{Soares-Furtado et al.}
\shorttitle{
K2 C0 Image Subtraction Reduced Light Curves
}
\ifthenelse{\boolean{emulateapj}}{
    
}{
    
}

\begin{document}
\title{
Image Subtraction Reduction of Open Clusters M35 \& NGC 2158 in the
\textit{K2} Campaign-0 Super-Stamp
}

\author{
    M.~Soares-Furtado\altaffilmark{1,$\dagger$},
    J.~D.~Hartman\altaffilmark{1},
    G.~\'A.~Bakos\altaffilmark{1,*,**},
    C.~X.~Huang\altaffilmark{2},
    K.~Penev\altaffilmark{1},
    W.~Bhatti\altaffilmark{1}
}

\altaffiltext{1}{Department of Astrophysical Sciences, Princeton
  University, Princeton, NJ 08544, USA; email: msoares@astro.princeton.edu}
\altaffiltext{$\dagger$}{National Science Foundation Graduate Research Fellow}
\altaffiltext{$*$}{
Alfred P. Sloan Research Fellow
}
\altaffiltext{$**$}{
Packard Fellow
}
\altaffiltext{2}{Astrophysics Department, Dunlap Institute for
  Astronomy and Astrophysics, University of Toronto, Toronto,ON M5S 3H4,
  Canada}

\begin{abstract}
\setcounter{footnote}{10}

Observations were made of the open clusters M35 and NGC 2158 during the
initial \textit{K2} campaign (C0). Reducing these data to high-precision
photometric time-series is challenging due to the wide point spread
function (PSF) and the blending of stellar light in such dense regions.
We developed an image-subtraction-based \textit{K2} reduction pipeline
that is applicable to both crowded and sparse stellar fields.  We
applied our pipeline to the data-rich C0 \textit{K2} super-stamp,
containing the two open clusters, as well as to the neighboring postage
stamps.  In this paper, we present our image subtraction reduction
pipeline and demonstrate that this technique achieves ultra-high
photometric precision for sources in the C0 super-stamp.  We extract the
raw light curves of 3960 stars taken from the UCAC4 and EPIC catalogs
and de-trend them for systematic effects.  We compare our photometric
results with the prior reductions published in the literature. For
detrended, TFA-corrected sources in the 12--12.25 $\rm K_{p}$ magnitude
range, we achieve a best 6.5 hour window running rms of 35\,ppm, falling
to 100\,ppm for fainter stars in the 14--14.25 $ \rm K_{p}$ magnitude
range. For stars with $K_{p}> 14$, our detrended and 6.5 hour binned
light curves achieve the highest photometric precision.  Moreover, all
our TFA-corrected sources have higher precision on all timescales
investigated.  This work represents the first published image
subtraction analysis of a \textit{K2} super-stamp. This method will be
particularly useful for analyzing the Galactic bulge observations
carried out during {\em K2} campaign 9. The raw light curves and the
final results of our detrending processes are publicly available at
\url{http://k2.hatsurveys.org/archive/}.

\setcounter{footnote}{0}
\end{abstract}

\section{Introduction}
\label{sec:introduction}

Since its launch in April of 2008, the \textit{Kepler Space Telescope}
has systematically detected an unprecedented number of exoplanet
candidates from the photometric signatures that these sources impart as
they transit their host stars (e.g.\ \citet{Mullally:2015}).
Undoubtedly, the \textit{Kepler} mission has played a pivotal role in
the field of exoplanetary science, contributing the largest catalog of
exoplanet candidates to date: 2/3rd of the current exoplanetary
census data \citep{Morton2016}.

The \textit{Kepler} photometer is a single-purpose instrument with a
0.95-m aperture Schmidt telescope design and a wide,
$\sim$100-square-degree field of view (FOV).  A detailed description of
the \textit{Kepler} mission can be found in \citet{Borucki:2010} and
\citet{Koch:2010}. During the primary phase of the \textit{Kepler}
mission, the spacecraft pointed toward a single patch of sky,
simultaneously observing more than 100,000 stars.

In 2013, after four years of service, it was necessary to revise the
direction of the mission after the failure of the second gyroscopic
reaction wheel, a requisite to maintain telescope pointing stability.
This event ushered in the second phase of the mission, \textit{the K2
  Ecliptic Survey} (\textit{K2}), designed to exploit the solar
radiation and firing of thrusters as a means of maintaining pointing
precision \citep{Howell:2014}.  \textit{K2} operations started in June
of 2014.  Remarkably, aside from the failure of the two reaction wheels,
the \textit{Kepler} spacecraft exhibits little performance degradation
and fuel budget estimates suggest a duration of 2--3 years for this
second phase.

\textit{K2} observes a series of target fields, known as campaigns,
along the plane of the ecliptic, for a span of $\sim$75 days each.
Through the numerous pointings scanning a multitude of Galactic
coordinates, the \textit{K2} mission provides a novel opportunity to
probe transiting planets among diverse stellar populations.  Each
individual campaign targets $\sim$10,000--20,000 stars to be observed at
29-min cadence, as well as an additional $\sim$100 targets that are
observed at 1-min cadence.  Observations are also made for a number of
open and globular clusters, including M35, NGC 2158, M4, M80, M45, NGC
1647, the Hyades, M44, M67, and NGC 6717.  The data are made publicly
available in a series of data releases. To date, Campaigns 0--8 have
been publicly released and are available on NASA's Barbara A.~Mikulski
Archive for Space Telescopes (MAST).

Our campaign of interest, C0, is the first target field, observed during
March-May 2014, and pointed toward the dense Galactic anti-center.
Approximately 22,000 targets were observed in C0.  Additionally, the
open clusters M35 (NGC 2168) and NGC 2158 were observed during this
campaign in what is known as a \textit{super-stamp} ---a contiguous
aggregate of 154 separate postage stamps ($50 \times 50$ pixels each)
placed over the densest region of these neighboring clusters. The open
cluster M35 is at a distance of $762\pm145$\,pc and has an estimated age
of 150\,Myr \citep{McNamara:2011}. It is relatively sparse compared to
NGC 2158, which is at $3600\pm400$\,pc, and 2\,Gyr old
\citep{Carraro:2002}. NGC 2158 is very dense, and was once believed to
be a globular cluster.

Open clusters offer an invaluable opportunity to probe stellar and
planetary astrophysics given the availability of cluster parameter
constraints (e.g.\ age, metallicity, Galactic position and motion), as
well as parameters for stellar members (e.g.\ stellar mass and
evolutionary state).  Moreover, the age of M35 makes it an ideal
environment to study planetary evolution, as planets are known to
undergo rapid evolutionary changes during the first few hundred million
years after their formation \citep{Adams:2006}.  To date, only a handful
of candidate exoplanets have been found in open clusters, primarily
through radial velocity measurements (e.g.\ \citet{Sato:2007,
  Lovis:2007, Quinn:2012, Libralato:2016}).  Using the \textit{Kepler}
satellite, \citet{Meibom:2013} unveiled the first transiting exoplanet
detection in an open cluster. More recently, data from the \textit{K2}
mission has continued to reveal transiting exoplanets in open cluster
environments \citep{Libralato:2016,Mann:2016b, Mann:2016}.  There
remains much to be learned regarding this intriguing class of objects,
such as whether exoplanets are a ubiquitous presence in dense open
cluster environments.

In crowded open cluster fields, however, heavy blending of light from
neighboring stars is unavoidable.  This problem is only amplified by the
large pixel size, wide PSF, and the lower pointing stability of the
\textit{K2} mission.  As a result, simple aperture photometry is not an
optimal means of obtaining high-precision stellar photometry.  Our work
aims to fully exploit the data-rich \textit{K2} super-stamps using an
image subtraction reduction technique, also known as ``differential
image analysis,'' outlined by \citet{Alard:1998}.  We fully automate
this procedure into a reduction pipeline and apply it to sources in the
C0 super-stamp.

There have been numerous investigations focusing on the same data set in
the past.  \citet{Vanderburg:2014} derived photometry for C0 data using
an efficient detrending correction technique outlined in
\citet{vanderburg_johnson:2014}.  This work focused solely on proposed
\textit{Kepler} target postage stamps, omitting cluster members and
neighboring stars located on the C0 super-stamp.  The first variability
search on C0 target postage stamps was performed by
\citet{Armstrong:2015}, also excluding the C0 super-stamp.  This
culminated in the identification of 8395 variable sources.

The first photometric results for C0 cluster members were obtained by
(\citet{Libralato:2016a}, hereafter L16), curtailing the averse effects of
light blending by employing a well-established PSF neighbor-subtraction
method, known to be an effective tool for extracting high-quality time
series data in dense fields, and then performing aperture photometry
\citep{Montalto:2007}.  They employ the high-angular-resolution Asiago
Input Catalog (AIC), assembled from observations by the ground-based
Asiago Schmidt Telescope.  This catalog lists 75,935 objects in the
region containing M35 and NGC 2158 \citep{Nardiello:2015}.  While the
AIC extends to faint magnitudes of $K_{p}\simeq24$, the rms scatter
(defined as the $3.5 \rm \ \sigma$-clipped 68.27th-percentile of the
distribution about the median value of the light curve magnitude) for
such dim stars is on the order of $\rm 50$ times greater than that of
sources in the 10--11 magnitude range. Moreover, as our primary aim is
to search for planetary transits in dense \textit{K2} fields, we focus
our sources for which reasonable photometry can be retrieved ($\rm rms <
0.02$).

The PSF neighbor-subtraction method requires high accuracy in the PSF
modeling, otherwise the technique will produce false residuals and
systematic errors --- a concern that our image subtraction technique
circumvents. The work of L16 resulted in a catalog
of 2,133 variable stars found within the C0 super-stamp.

Our work provides the first image subtraction reduction of the
C0 super-stamp, effectively removing sources with no detectable
variability from the cluster field, therefore reducing blending, to
produce high-precision differential photometry.  While applying the
image subtraction reduction technique to \textit{K2} super-stamps is
novel, the method itself is not a new concept.  \citet{Alard:1998}
outlined this procedure nearly two decades ago, releasing the
\texttt{ISIS} package and then further optimizing this process by
incorporating a space-varying convolution kernel \citep{Alard:2000}.  We
make use of the image subtraction implementation of the HATNet project
\citep{Bakos:2010} as described by \citet{Pal:2012}.

The crowding from variable sources and the photon-noise residuals of
non-variable sources on the image-subtracted frames is much smaller than
crowding present on the original unsubtracted frames.  This is because
the vast majority of photometric sources tend to be either not
detectably variable or they are only variable over long timescales.  The
image subtraction technique therefore offers the major advantage of far
less blending in the resulting photometry.  Furthermore, rather than
modeling each of the images for a given cadence, we model the {\em
  changes} between images, which include variations in pointing, flux
scaling, background, and the convolution kernel relating PSFs.  These
variations tend to be simpler to accurately model and are generally well
fit by simple functions.  The PSF, background, star positions, and
relative fluxes are determined only once for a single photometric
reference frame, which, in our case, is the median valued co-added frame
taken from the entirety of our selected data set.  Moreover, systematic
errors that arise produce an increase in the overall amplitude of a
light curve (a scaling error), rather than contributing to light curve
noise.  In contrast, proper modeling of the non-subtracted frames
requires accurate modeling of the PSF, positions, background, and
relative fluxes, which is far more challenging, particularly for an
open cluster region where blending is profuse.  One additional advantage
of the image subtraction is that the source of the variation is often
uniquely identified from the excess (or missing) residual flux at a
given location, even under strong crowding.

In this paper, we reduce data from the \textit{K2} C0 super-stamps and
neighboring target pixel files to produce high-precision photometry for
sources down to $K_{p}\sim16$, employing techniques developed from the
HAT ground-based transit surveys \citep{Bakos:2010} and building upon
the work of \citet{Huang:2015}.  We publicly release the raw and
detrended high-precision light curves.  We briefly review K2
observations and describe C0 data extraction in Section~\ref{sec:obs}.
The data reduction method is reviewed in Section~\ref{sec:reduction},
including astrometry, image subtraction, and photometry.  Our detrending
techniques are discussed in Section~\ref{sec_detrend}.  Finally, in
Section~\ref{sec:discussion}, we compare our C0 photometry and light
curves with those obtained from prior studies (specifically that of L16),
and demonstrate that we have generated the most precise photometric
analysis of sources in C0 super-stamp.  This is followed by a summary of
our current efforts to conduct a variability search of our time series
data.

\section{Observations}
\label{sec:obs}
\subsection{\textit{K2} Data Acquisition}
\label{subsec:acq}

A thorough review of the \textit{Kepler} instrument is given in
\citetalias{KeplerHandbook:2009}.  Here we summarize the principal
features of \textit{K2} data acquisition.

The \textit{K2} photometer is comprised of a 21-module array covering
5-square-degrees on the sky, providing $\sim$100-square-degree FOV.
Each module contains two separate $\rm 2200 \times 1024$ pixel CCDs for
a total of $\sim$95 million pixels across the array.  Two of the 21
modules were not operable when these observations were made and, more
recently, a third module has also failed.  Each module contains 4 output
channels, designated by channel numbers 1--84.  To prevent saturation,
CCDs are read out every six seconds and the data are integrated for either
a \textit{long} 29-minute cadence or a 1-minute \textit{short} cadence.
To improve the photometric precision, images are de-focused to produce
10 arc-second wide PSFs.

In the \textit{K2} phase the spacecraft is pointed toward the ecliptic
in order to minimize the impact of solar radiation pressure.  The still
functioning two reaction wheels respond to the pressure exerted by solar
radiation, providing a close to constant pointing alignment in the y and
z axes, while the thrusters are fired to correct for drift along the
x-axis every $\sim 6$ hours.

The coordinates of target sources in the \textit{K2} mission are
provided by the Ecliptic Plane Input Catalog (EPIC).  The number of
target objects is limited by the onboard storage, compression, and
downlink capabilities, as well as the duration of the campaign.
Observations of each target source, also known as a \textit{Kepler
  Object of Interest} or \textit{KOI}, are downloaded once per month as
a $\rm 25 \times 25$ pixel postage stamp centered on the target,
comprising $10 \%$ of the entire \textit{Kepler} field --- although some
postage stamps can be as large as 50-pixels across.  super-stamps are
assigned a custom aperture number to serve as an identifier.  Also
obtained are two Full Frame Images (FFI) at the start and end of each
campaign.

\subsection{\textit{K2} C0 Data Extraction}
\label{subsec:data}
Our region of interest is the C0 super-stamp containing the open
clusters M35 and NGC 2158, comprised of 385,000 individual pixels.
These data are found on a single module output channel --- channel 81.
We make use of the \textit{long}, 29-min cadence observations.  Our data
were obtained as target pixel files (TPFs) from NASA's Barbara
A.~Mikulski Archive for Space Telescopes (MAST). TPFs are the time
series pixel data for a particular stamp, centered on a target object.
Unfortunately, the first half of the C0 observations were conducted in
coarse pointing mode, resulting in large positional variations by up to
$\sim$20 pixels ($25 \ \rm mas \ pixel^{-1}$).  Therefore, we solely
analyze the second half of the data set, which has a baseline of
${\sim}31$ days (1840 cadences), where fine pointing mode was employed
and the positional variations are significantly diminished.  In
contrast, the light curves generated by L16 employ the full data set.

\section{Data Reduction}
\label{sec:reduction}
The data reduction process employed for our image subtraction pipeline
is based on the procedures described by \citet{Huang:2015}.  Here we
give a brief outline, and then discuss the process in more detail:
\begin{enumerate}
\item For each cadence, we employ header information from the TPFs
  to generate a sparse frame image, which contains all sources observed
  on the \textit{K2} super-stamp and adjacent stamps.
\item We perform source extraction on the stars present in the
  \textit{K2} super-stamp and adjacent TPFs and find an absolute
  astrometric transformation between the UCAC4 catalog (RA and Dec)
  and the extracted (x, y) positions of the sources in the frame.
\item An astrometric reference frame is selected. We use a sharp sparse
  frame image with median directional pointing.
\item We then spatially transform all sparse frame images to a common
  astrometric reference coordinate frame. This is accomplished by
  finding a transformation between the pixel coordinates of source
  centroids in a particular sparse frame image and the selected
  astrometric reference frame. This is a crucial step, minimizing the
  effect of spacecraft drift and allowing for more accurate modeling of
  the motion of the instrument in our detrending analysis.
\item We generate a master photometric reference frame, which is a
  stacked median average of all available C0 frames, after transforming
  to a common astrometric frame and matching their backgrounds and PSFs.
\item Each sparse frame cadence file is matched to and then subtracted
  from the photometric reference frame, leaving behind subtracted images
  showing only the variability in the field.
\item We perform photometry on the catalog projected sources (using
    the absolute astrometry above).
\item We then assemble light curves for all of the sources.
\item We apply a high-pass filter to our data (using 1-day binning) and
  perform a parameter de-correlation procedure, analogous to that
  outlined by \citet{vanderburg_johnson:2014}, to remove the remaining
  small, time-correlated
  noise from our light curves.
\item We then apply the \textit{Trend Filtering Algorithm}.
  \citep{Kovacs:2005}, as implemented in {\sc vartools}
  \citep{Hartman:2016} to further remove systematic artifacts from the
  data.
\end{enumerate}
  
\subsection{Source Catalog Preparation}
\label{subsec:cat} 
Our astrometry procedure is dependent upon an independent knowledge of
the true positions of the observed celestial sources.  For this, we
employ the fourth edition of the United States Naval Observatory CCD
Astrograph Catalog (UCAC4) \citep{Zacharias:2013}. This catalog is based
upon observations with a much higher spatial resolution than the
\textit{K2} observations and, as such, provides more accurate positions
and photometric measurements than what can be deduced from the data
alone.  UCAC4 covers a stellar brightness range of 8th-16th magnitude in
a single bandpass between V and R.  Positional errors are on the order
of 15--20 mas for stars in the range 10th-14th magnitude.  Measurements
are based on the International Celestial Reference System (ICRS) at a
mean epoch of 2000.  In addition to positional coordinates, the UCAC4
lists measurements of proper motion for ${\sim}92\%$ of cataloged stars
with errors on the order of 1--10 $\rm mas \ yr^{-1}$.  In the work
performed by \citet{Huang:2015}, first-order corrections were applied to
the coordinates of catalog objects based upon the proper motions to
account for positional changes.

These catalogs are imperative to determining accurate brightness
measurements for sources, as they have high spatial resolution, 
and are therefore less impacted by blending.  The UCAC4 is supplemented
by photometric data from the Two Micron All Sky Survey (2MASS)
\citep{Skrutskie:2006} for ${\sim}92\%$ of cataloged stars (J, H and K
bands), as well from the AAVSO Photometric All-Sky Survey (APASS) for
$\sim 45\%$ of objects (BVgri bands).

The \textit{Kepler} magnitude ($K_{p}$), a measure of the source
intensity as observed through the wide \textit{Kepler} bandpass, is
computed for all sources using pre-existing photometric data.  Following
the hierarchical conversion method outlined in \citet{Brown:2011}, we
employ \textit{gri} band photometry whenever possible.  For sources
without these measurements, however, we employ \textit{BV} band
magnitudes from the Tycho-2 catalog \citep{Tycho:2000}.  For a subset of
sources, we use B and V magnitude measurements to estimate
\textit{Kepler} magnitudes.

In addition to the \textit{K2} target sources, our astrometric analysis
provides positions and photometry for many other stars falling on the
super-stamp and adjacent target pixel files. In total, 120 of our final
light curves are K2 target sources, while the remaining 3840 light
curves are for objects that are not targeted K2 sources.

\subsection{Image Preparation and Astrometry}
\label{sec:astrom}

\subsubsection{Sparse Frame Image Construction}
We make use of the publicly available \textit{K2} C0 TPFs, generating a
sparse frame image of the entire channel for each cadence by stitching
together the individual target pixel stamps.  This is a useful
alternative to working with a multitude of separate target pixel
stamps. We use the available fits header information to determine the
proper placement of target pixel stamps and compare our results with the
FFI for that same field.

Each sparse frame image is comprised of a field of 1132 $\times$ 1070
square pixels, and unobserved regions in the sparse frame image are
masked. A \texttt{Python} script automates the translation and stitching
procedure, reconstructing a total of 1,392 high-quality sparse frame
images in fine-pointing mode, with each image corresponding to an
individual cadence.  These frames are cross-checked with the FFI to
ensure proper reconstruction.  Figure~\ref{fig:stitchedframe} shows one
example of a sparse frame image.

\subsubsection{Relative Astrometry}

The next step is a translation of each of the cadence frames into a
common reference coordinate system.  The astrometric reference file is a
sparse frame image that sets the coordinate reference frame. The
astrometric reference file is selected for its sharp stellar profiles,
as compared to the other files, and a pointing position that is close to
the median pointing over the duration of the observations.

We perform source extraction on the astrometric reference frame.  We use
the tool \texttt{fistar}, available in the astronomical image and data
processing open-source software package \texttt{FITSH}, to detect and
extract sources from the images. The \texttt{fistar} tool designates
source candidate pixel groups, modeled in our case by an asymmetric
Gaussian profile in order to derive precise centroid coordinates and
shape parameters. We then extract the PSF for the image, based on the
modeled sources. The output file is a list of detected and extracted
sources (e.g.\ positions, S/N, Flux).

We then employ the \texttt{FITSH} tool \texttt{grmatch} to match
detected sources from each of the sparse frame images to that of the
astrometric reference frame. Matching is performed using a 2-dimensional
point matching algorithm, determining an appropriate geometrical
transformation to transform the points from each sparse frame cadence
image to the astrometric reference frame, finding as many pairs as
possible \citep{Pal:2006}.

The output file contains the relevant geometrical transformation and
statistics regarding the quality of the transformation.  We employ an
adapted version of the \texttt{FITSH} tool \texttt{fitrans} to perform
the appropriate geometric transformations on each of the sparse frame
cadence files. We use linear interpolation between the pixels involving
exact flux conservation by integrating on the image surface.

The relative astrometry step is vital as it (a) generates accurate
centroids for sources and (b) mitigates negative effects introduced by
the spacecraft roll, which allows for more accurate modeling of the
instrument motion in our detrending procedure.
The original \textit{Kepler} mission was known for its excellent
pointing stability, while \textit{K2} observations are plagued by
significant pixel drift, with a typical star shifting along a
${\sim}$2-pixel long arc in our C0 data set.  Variations in the pixel
sensitivity produce fluctuations in the measured flux as a function of
centroid position.  Not surprisingly, the resulting PSF is distorted and
blending is recurrently inevitable, particularly in dense stellar
regions.  Determining an accurate PSF shape is a crucial step in
obtaining high-precision photometry in crowded fields like C0.

\subsection{Image Subtraction}
\label{sec:imagesub}

For \textit{Kepler} and \textit{K2}, aperture photometry is the primary
method employed to derive light curves.  Certain regions, however, are
crowded, partly due to \textit{Kepler's} large, undersampled pixels
($\sim$4 arcsec/pixel). Moreover, the PSF extends across several
pixels. Here aperture photometry can perform sub-optimally.  Subtracting
out the flux from non-varying sources is a means of teasing out the
photometric variable sources in crowded regions L16 used the
high-resolution AIC (Asiago Input Catalog) to identify neighbors for
sources in the C0 super-stamp field, subtracting out the contributions
from these contaminants.

Our method, however, instead relies on the creation of a stacked frame,
coined the \textit{photo-reference frame}, comprised of all 2419
\textit{K2} C0 frames.  We explored multiple methods to generate the
optimal photo-reference frame by varying the number of frames in the
stack, employing quality cuts on frames included in the stack, and
trying different averaging techniques. After comparing the resulting
light curve quality and rms scatter of these routes, we conclude that
the optimal photo-reference frame is generated by astrometrically
matching our sparse frame images with \texttt{fitrans}, convolving the
set of frames to match their corresponding PSFs and backgrounds, and
then combining them to take the mean flux.  We use the \texttt{ficonv}
tool from the FITSH package to match and convolve the frames. We note
that matching the backgrounds is particularly important due to the
considerable changes in the zodiacal background light over the campaign.

After generating the photo-reference frame, we then use \texttt{ficonv}
to match the background and PSF of each stitched frame cadence image to
that of the photo-reference image before subtracting the two frames from
one another.  In order to properly do this we optimize the selected
convolution kernel parameters, settling on a discrete image subtraction
kernel over a Gaussian kernel.

\begin{figure*}
  \centering
  \includegraphics[width=0.7\textwidth]{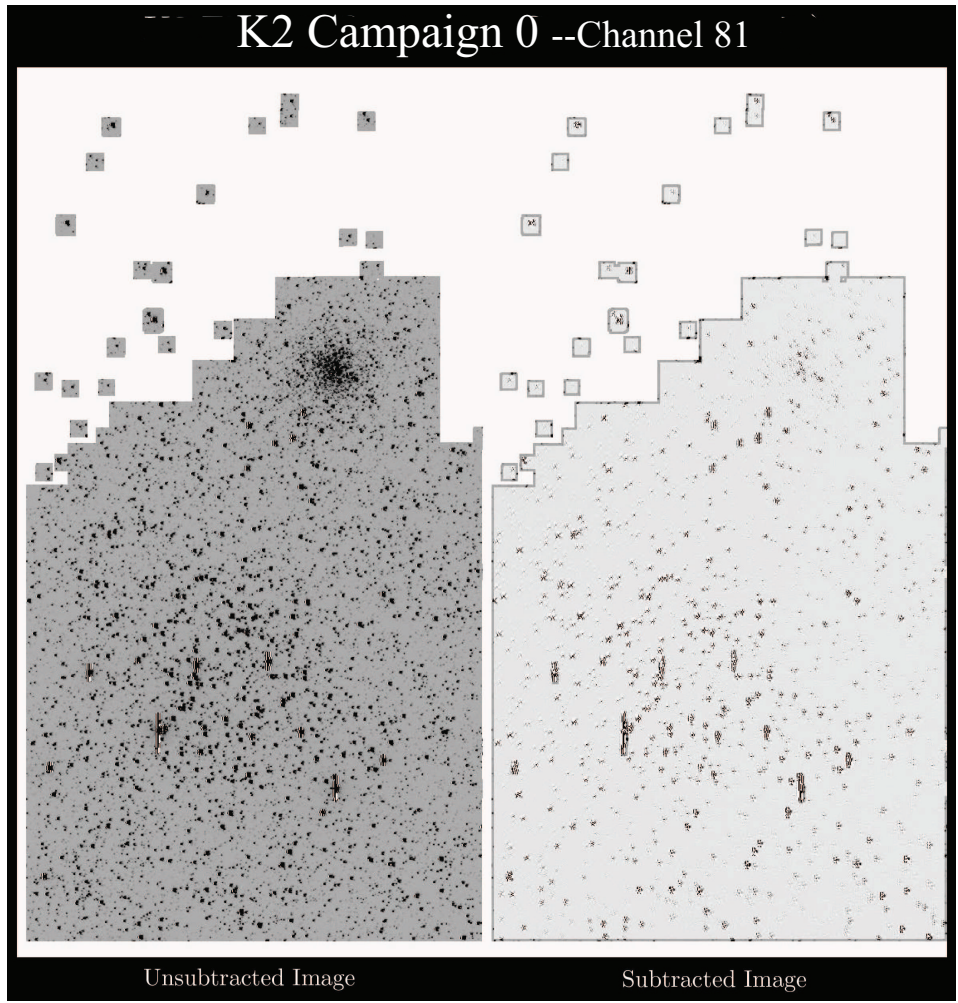}
  \caption{$\textit{K2}$ Campaign 0 super-stamp containing open clusters
    M35 and NGC 2158. The figure is generated by stitching stamps within a
    given channel (channel 81 in this case) into a composite image. The
    empty regions represent null pixels, as the information from these
    pixels are not downloaded by the spacecraft. The left panel displays
    the raw composite frame before our image subtraction reduction
    pipeline is applied. The right panel is an example of a single
    subtracted frame for an arbitrary cadence. Variable sources remain in
    the subtracted frame as well as bright stars that are still visible
    due to saturation or Poisson noise.}
  \label{fig:stitchedframe}
\end{figure*}

With the frames appropriately convolved, we subtract photo-reference
frame from each of the convolved individual cadences, leaving behind a
sparser stellar field of varying sources (and residuals left after
constant stars), as shown in Figure~\ref{fig:stitchedframe}.

In addition to reducing blending in crowded regions, image subtraction
affords the ability to uniquely identify sources from the excess (or
missing) residual flux at a given location, even in dense stellar fields.

\subsubsection{Image Subtraction Kernel}

We parameterize the transformation that matches an observed image to the
photo-reference image using a model of the form:
\begin{eqnarray*}
  I^{\prime}_{ij} & = & \sum_{l=0}^{l=O_{B}-m}\sum_{m=0}^{m=O_{B}}b_{lm}i^{l}j^{m} \nonumber \\
 & & + \sum_{l=0}^{l=O_{I}-m}\sum_{m=0}^{m=O_{I}}K_{00lm}i^{l}j^{m} \times I_{ij} \nonumber \\
 & & + \sum_{l=0}^{l=O_{K}-m}\sum_{m=0}^{m=O_{K}}\sum_{i^{\prime}, j^{\prime} =
    -k, i^{\prime}j^{\prime} \neq 00}^{i^{\prime}, j^{\prime} = k} K_{i^{\prime}j^{\prime}lm}i^{l}j^{m} \\
 & & \times \ (I_{i + i^{\prime}, j+j^{\prime}} - I_{ij})/2,
\end{eqnarray*}
where $I_{ij}$ represents the image undergoing analysis,
$I^{\prime}_{ij}$ is the corresponding transformed image.  The free
parameters $b_{lm}$ signify changes in the background, $K_{00lm}$
signify changes in the flux scaling, and $K_{i^{\prime}j^{\prime}lm}$,
with $i^{\prime}j^{\prime} \neq 00$, represent the discrete convolution
kernel modeling changes in the shape of the PSF.  We use polynomials in
the spatial coordinates $(i,j)$ to represent spatial variations in each
of these transformations. The order of the polynomial is given by
$O_{B}$, $O_{I}$ and $O_{K}$ for the background, flux scaling, and PSF
shape changes, respectively. The free parameters are linearly optimized
to minimize the sum
\begin{equation*}
\sum_{ij} (I^{\prime}_{ij} - R_{ij})^2,
\end{equation*}
where $R$ represents the photo-reference image.  We explored a variety
of values for the half-size of the discrete convolution kernel ($k$),
and the spatial polynomial orders $O_{B}$, $O_{I}$ and $O_{K}$. We
settled on the optimal value of $k=2$. We also found that assuming no
spatial variation in the parameters ($O_{B} = O_{I} = O_{K} = 0$)
\emph{generally} provided the best results.

The optimal kernel parameters for each source are listed in the final
column in Table~\ref{table:tableNoTFA} in the format ``$\rm b0i0d20$''
(the numbers here represent example discrete kernel values that produces
high-quality results).  The number following `b' represents the optimal
value of $O_{B}$ for that source.  The value following `i' sets $O_{I}$.
The two numbers following `d' represent $k$ and $O_{K}$,
respectively. For all sources we find that the light curve scatter is
minimized when adopting $O_{I} = O_{K} = 0$ and $k = 2$. The optimal
value for the order of the constant offset background kernel $O_{B}$,
however, varies on a source-by-source basis, ranging between 0--4.

\subsection{Photometry}
\label{sec:photometry}

There are two steps in performing photometry using subtracted images:
1.\ measuring the total fluxes of each of the sources on the
photo-reference image, and 2.\ measuring the differential fluxes of each
of the sources on each of the subtracted images. The total flux of a
given source on a given image is then the sum of the reference flux of
the source and the differential flux of the source for the image in
question.

Determining the reference fluxes in this case is complicated due to the
low spatial resolution of K2 and the significant blending of sources in
the field of these clusters. We therefore make use of the UCAC4
determined K2 magnitudes, which are based on higher spatial resolution
images, to set the relative fluxes of all sources in the field. To
determine the zero-point flux level for the photo-reference image, we
perform aperture photometry on the image using the \texttt{fiphot} tool
in the \texttt{FITSH} package, and then determine the median ratio of
the aperture photometry flux to the UCAC4 determined flux for unblended
sources. Note that we are assuming in this process that none of the
sources have varied in brightness between the UCAC4 and the
photo-reference image, any violation of this assumption will lead to an
error in the amplitude of variations in our image subtraction light
curve, but will not distort the signal or add noise to it.

Differential fluxes are also computed using \texttt{fiphot}. We perform
aperture photometry on the subtracted images at the fixed locations of
the UCAC4 sources transformed to the image coordinates based on our
astrometric solutions determined in Section~\ref{sec:astrom}. In order
to ensure that the differential photometry is measured on a consistent
system, \texttt{fiphot} makes use of the kernel parameters determined by
\texttt{ficonv} to account for any changes in the PSF shape or flux
scale between the photo-reference and the subtracted image. We employ 9
apertures to measure the residual (differential) flux, 
ranging in size from 2.5 pixels to
5.0 pixels. The suggested optimal aperture for each source is then
determined on a source-by-source basis, selecting the aperture
with the lowest rms for each light curve.

\section{Light Curve Detrending}
\label{sec_detrend}
  
While the image subtraction procedure should in principle correct for
systematic variations in the images, leaving clean light curves free of
instrumental variations, we found that, in practice, systematic
variations remain, and post-processing is needed to remove these
variations.  As described in Section~\ref{subsec:acq}, the low frequency
roll of the satellite causes sources to drift across the FOV.  Thrusters
are fired on timescales of $\sim 6$ hours to correct for this
effect. This drift results in decreased photometric precision, as the
star dithers between pixels changes in the pixel sensitivity must be
taken into account. Fortunately this drift occurs largely along a single
axis in a well defined motion, which greatly simplifies the necessary
corrections.

There are several methods that have been employed to correct for
systematic effects in \textit{K2} data. These include a Gaussian process
detrending approach outlined by \citet{Aigrain:2015} as well as a method
discussed in \citet{FM:2015}, which performs simultaneous fit for
systematics.  We follow a procedure similar to the self flat-fielding
detrending method outlined by \citet{vanderburg_johnson:2014}, coined
`K2SFF', which corrects for drift systematics by decorrelating
photometric light curves with the spacecraft motion.

The first step of our automated detrending procedure is a culling
process whereby we remove all low-quality cadences. In the future, we
plan to apply weights to each cadence rather than simply omitting a file
from the data set. Our detrending method differs from that of
\citet{vanderburg_johnson:2014} in that we do not rely on thruster flags
labels as an indicator of file quality. Instead, we calculate the
cadence-by-cadence drift between source centroids, clipping cadences
where the drift exceeds 3.5 $\rm \sigma$. This method has an $82 \%$
overlap with thruster flagged cadences.

From this point, a median high-pass filter is applied to the culled data
with binning windows of $\sim 1$ day.  Principal component analysis is
then performed on the 2-dimensional scatter traced out by the source
centroids as they drift with the spacecraft roll.  This scatter is
primarily 1-dimensional, and we therefore solely utilize the predominant
basis vector. Corrections to account for changes in pixel response as a
function of arclength offset (calculated as per
\citealp{vanderburg_johnson:2014}) are applied.  To apply the
correction, we break up the predominant basis vector into 20 equally
sized arclength offset bins, determining the median flux for each
arclength bin. We then fit a B-spline function to the binned arclength
offset vs.\ median flux curve.

The results of our detrending procedure can be seen in the light curves
shown in Figure~\ref{plotcomparison}.  This figure is discussed in
greater detail in Section~\ref{sec:discussion}.

\subsection{Trend Filtering Algorithm}

In addition to the detrending process outlined above, we also include an
optional post-processing step with the application of the Trend
Filtering Algorithm (TFA), which is explained in detail in
\citet{Kovacs:2005}.  We employ 141 TFA template stars, which are
randomly chosen detrended light curve files containing more than $99\%$
of all cadences.

The TFA technique, when applied in non-reconstructive mode as we do
here, may not work well for long-period and/or high S/N variables.
Instead of filtering out purely instrumental systematic variations or
uncorrelated noise, the TFA will suppress and distort real variability
from the source.  For these high S/N and long period variables the
non-TFA detrended light curves are optimal.  The TFA does optimize our
results for low S/N variables --- where the signal is lower
than the effect produced by instrumental systematics, as in this case
the instrumental noise will be suppressed and the signal is more easily
identified.  If the magnitude variation is dominated by white noise, the
TFA has negligible impact on the resulting light curve.

\section{Results and Discussion}
\label{sec:discussion}

In this paper we have presented our image subtraction technique to
generate high-precision, detrended light curves for stars in the open
clusters M35 and NGC 2158, observed in the Campaign 0 \textit{K2} field.

As L16 has already addressed a direct comparison between the
neighbor-subtracted light curves and those generated in
\citet{Vanderburg:2014}, and since \citet{Vanderburg:2014} ignores the
crowded super-stamp, we compare our results only with L16 to determine
the robustness of the method.

\begin{figure*}
  \centering \includegraphics[width=0.6\textwidth]{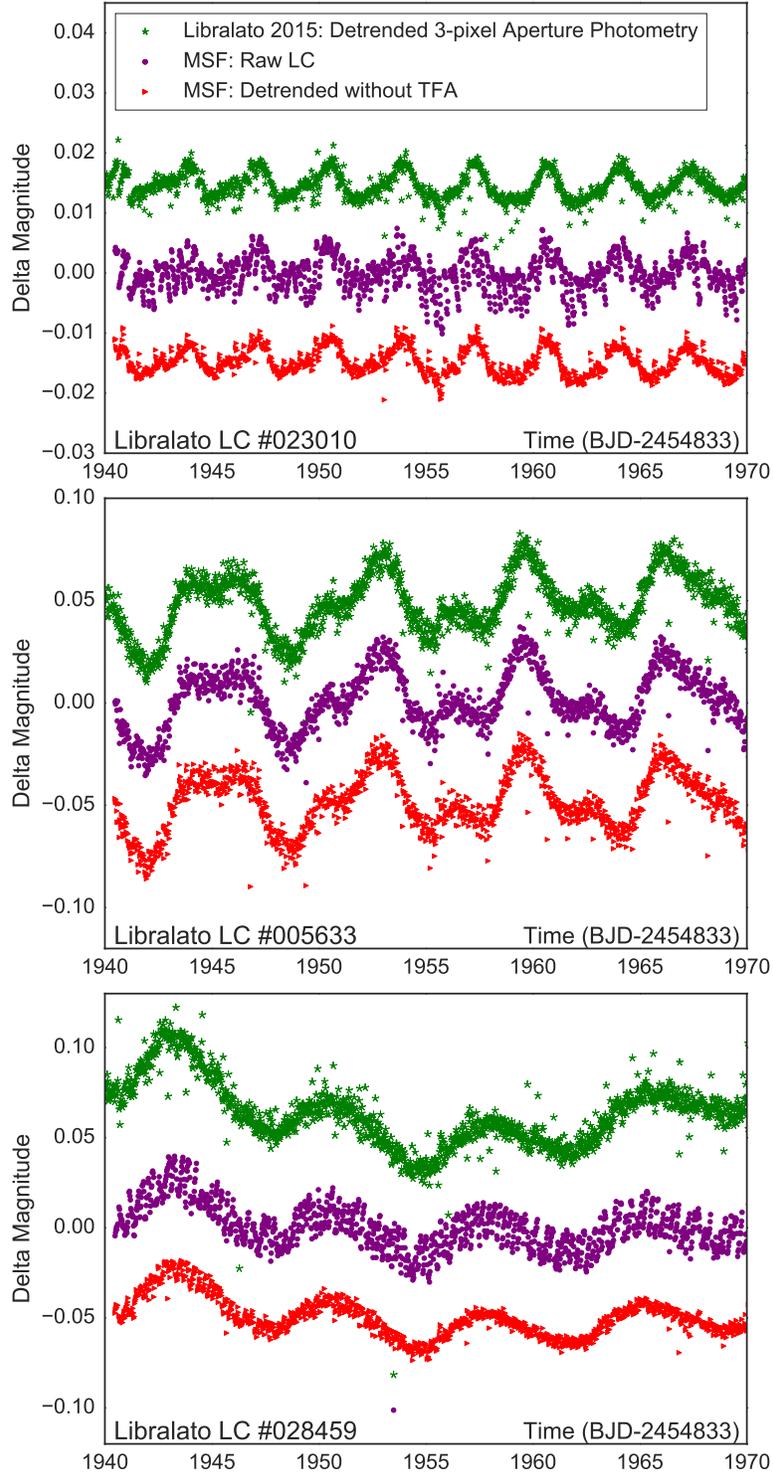}
  \caption{Direct comparison of our photometric results with known
    variable sources presented in L16. The \textit{green, star-shaped
      points} illustrate the best aperture detrended results produced by
    L16 (these happen to be the detrended three-pixel aperture photometric
    results), while the \textit{purple, round points} show our raw,
    image-subtracted, best aperture results for the same source. The
    \textit{red, triangle-shaped points} display our detrended,
    best aperture photometric results without application of the TFA
    procedure.  In both the top and middle panels, our results are
    comparable to that of L16, while in the bottom panel both our raw
    and detrended light curve show far less scatter.}
\label{plotcomparison}
\end{figure*}

Figure~\ref{plotcomparison} compares our light curves to those of L16
for three representative variable sources identified by the L16 variable
search. In general, the image subtraction light curves are comparable to
those generated by L16 using neighbor subtraction, although in selected
cases our precision is superior.

\begin{figure*}
  \centering
  \includegraphics[width=0.9\textwidth]{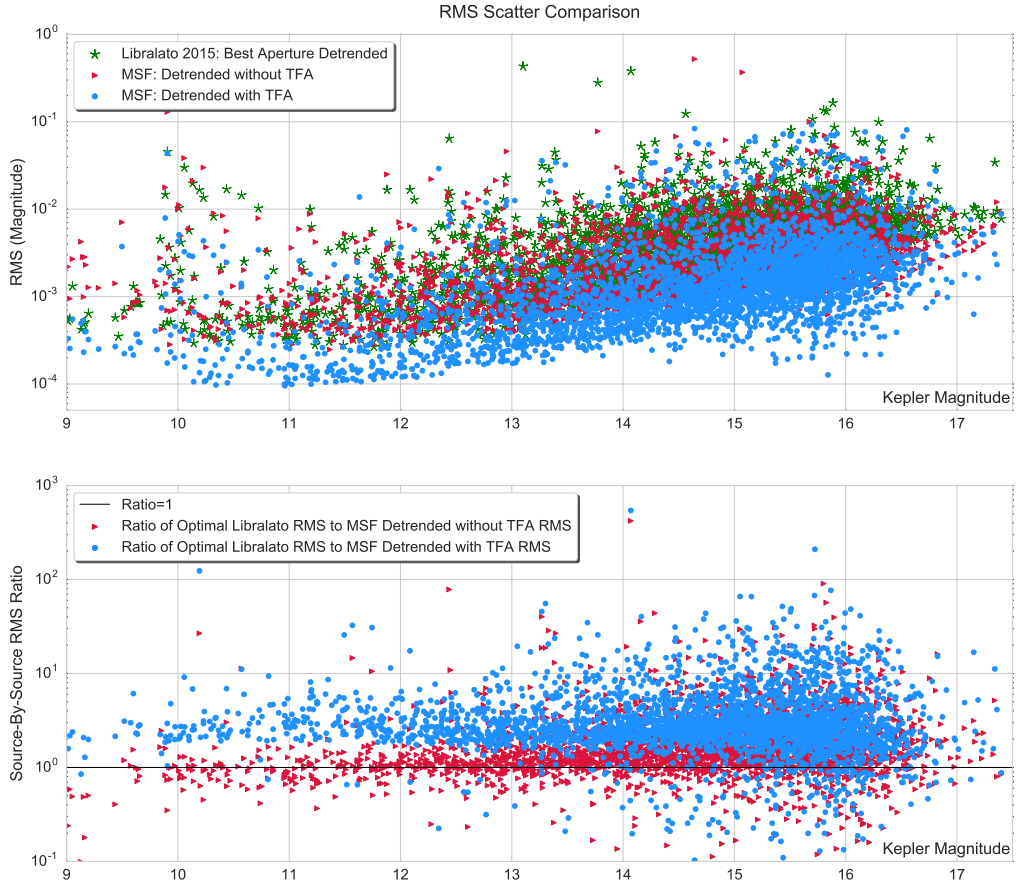}
  \vspace*{-10mm}
  \caption{Direct comparison of the rms magnitude scatter of our
    detrended results to the best results produced by L16.  In the top
    panel, we plot the rms magnitude versus Kepler magnitude for
    matching sources.
    L16 produced light curves for
    many more objects, especially faint sources with $K_{p} > 17$, but
    to provide a fair comparison of the precision of the two methods, we
    are only showing objects in these plots for which we also produced a
    light curve.
    The \textit{green, star-shaped points} depict the L16
    sources, while the \textit{red, triangle-shaped points} display the
    rms magnitude scatter for our detrended light curves without the TFA
    application. The \textit{blue, round points} show our rms magnitude
    scatter for TFA-corrected light curves. It is clear from the top
    panel that the TFA application reduces the rms considerably,
    however, not at a cost to the signal amplitude.
    Faint sources ($K_{p} \ga 15$) with anomalously
    low rms are sources for which the catalog Kepler magnitude is
    brighter than the true magnitude on the photo-reference image,
    leading to underestimating the amplitude of photometric variations
    when computed in magnitudes.
    In the bottom panel we plot the
    source-by-source ratio of the L16 optimal rms magnitude scatter to
    that of our detrended rms (\textit{red, triangle-shaped points}) and
    to that of TFA-corrected results (\textit{blue, dashed
      points}). For our detrended sources, $67.6\%$ lie above the
    \textit{black line} where $\rm ratio=1$, and for our post-TFA
    sources $76.6\%$ lie above this line.}
  \label{fig:magrms}
  \end{figure*}

In addition to comparing individual sources, the top panel of
Figure~\ref{fig:magrms} shows the light curve rms scatter (at cadence)
vs.\ {\em Kepler} magnitude for our pre-TFA detrended light curves, our
post-TFA light curves, and the best aperture light curves from L16, to
provide the reader with a sense of how these methods compare.  The
bottom panel shows the source-by-source ratio of the rms scatter of the
L16 derived light curves our light curves as a function of magnitude.
For our detrended sources, $67.6\%$ lie above the black $\rm Ratio=1$
line, and for our post-TFA sources $76.6\%$ lie above this line,
indicating a reduction in the light curve rms, which does not come at a
cost to the signal amplitude.

\begin{figure*}
  \centering \includegraphics[width=0.8\textwidth]{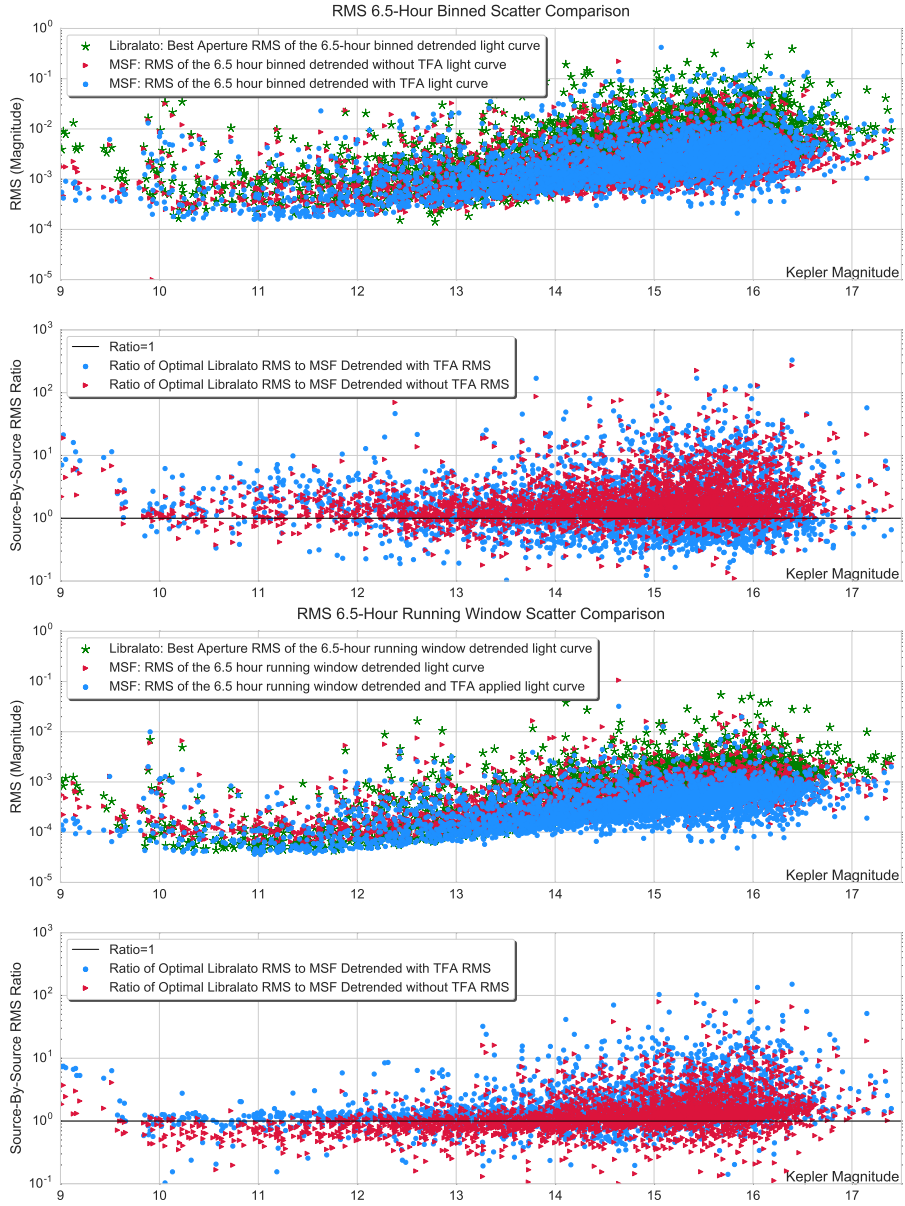}
  \vspace*{-20mm}
  \caption{Direct comparison of the 6.5 hour binned rms magnitude
    scatter of our detrended and TFA-corrected results to the best L16
    results in the top panel.  The \textit{green, star-shaped points}
    depict the 6.5 hour binned rms of the magnitude for all the L16
    sources. The \textit{red, triangle-shaped points} display our 6.5
    binned rms magnitude scatter for detrended results without TFA,
    while the \textit{blue, circle-shaped points} illustrate the
    magnitude scatter for results that are TFA-corrected. For a fair
    comparison, we do not show all L16 sources, and instead display only
    source matches between both data sets. The second panel from the top
    shows the source-by-source ratio of the L16 rms to our results for
    both the detrended results (\textit{red, triangle-shaped points})
    and the TFA-corrected results (\textit{blue, round points}). The
    \textit{solid black line} displays where $\rm ratio=1$. The third
    panel illustrates a comparison of the 6.5 hour running window rms
    magnitude scatter of our detrended and TFA-corrected results to that
    of L16. Once again, \textit{green, star-shaped points} depict L16
    sources, \textit{red, triangle-shaped points} represent our
    detrended sources without TFA, and \textit{blue, circle-shaped
      points} depict our TFA-corrected sources.  In the bottom panel, we
    plot the source-by-source ratio of the L16 6.5 hour running window
    rms to our results for both the detrended results (\textit{red,
      triangle-shaped points}) and the TFA-corrected results
    (\textit{blue, round points}). The \textit{solid black line}
    displays where $\rm ratio=1$.  It is worth repeating that the
    reduced scatter in our light curves does not come at a cost of the
    signal amplitude.}
  \label{fig:Binned}
\end{figure*}

In the top panel of Figure~\ref{fig:Binned} we show the same results
after binning the light curves by 6.5\,hr, and in the third panel we
show the results when computing the 6.5\,hr running-window rms \ as
defined by \citet{Vanderburg:2014} (this is computed to allow a more
direct comparison to prior results, and is calculated by dividing the
long-cadence light curves into bins of 13 consecutive cadences,
computing the rms in each bin and dividing by $\sqrt{13}$, and then
taking the median value over all bins). The second and fourth panels of
Figure~\ref{fig:Binned} display the source-by-source ratios (L16 to our
data set) of the 6.5\,hr binned rms and 6.5\,hr running window rms
(using only quality flagged data), respectively.  We determine the
ratios for both our detrended results and the TFA-corrected results.

\begin{figure}
  \centering \includegraphics[width=0.5\textwidth]{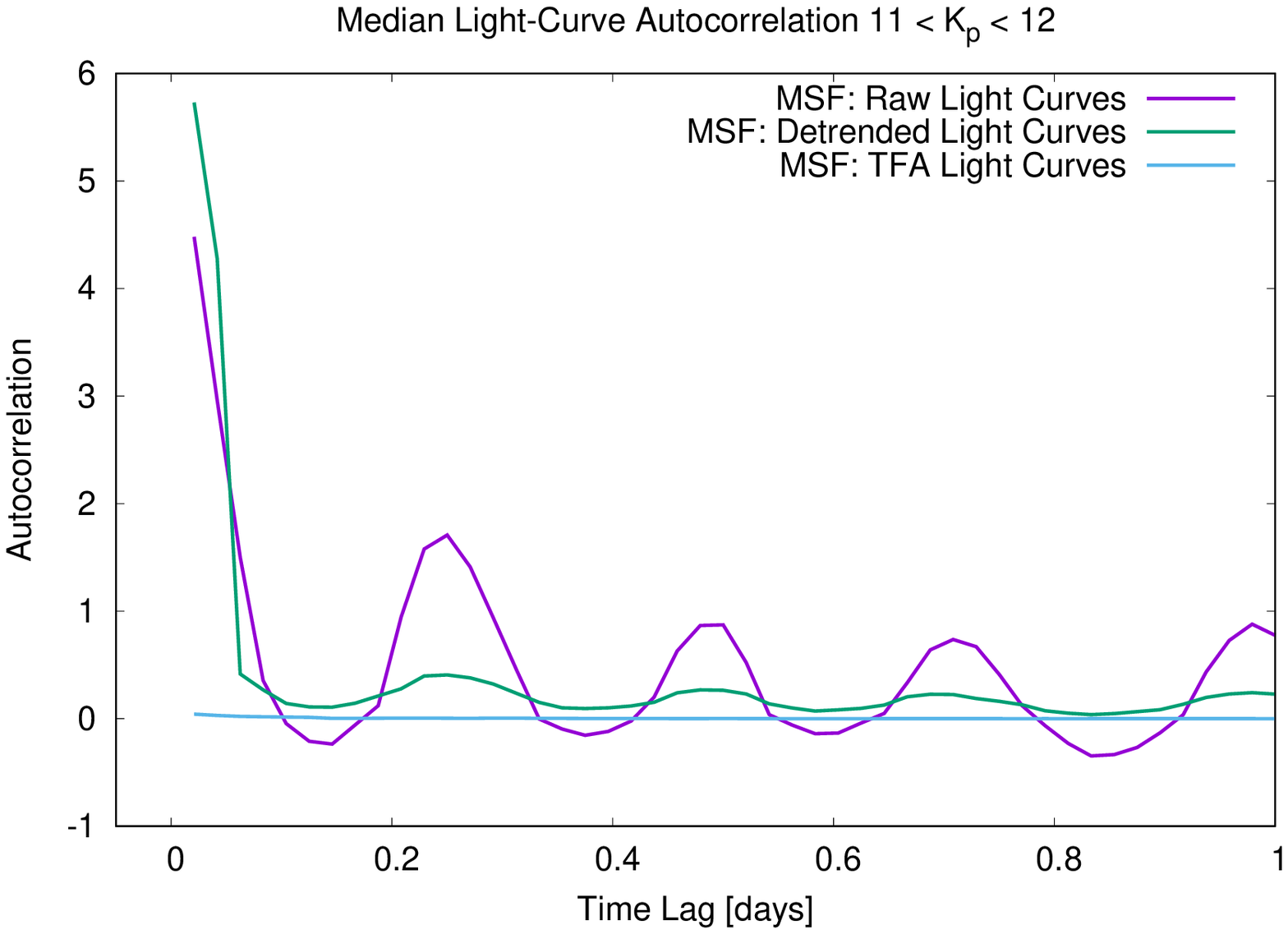}
  \caption{Median discrete autocorrelation function for sources
    with $11 < K_{p} < 12$, shown separately for our raw light
    curves, detrended light curves, and TFA-corrected light
    curves. The autocorrelation is computed relative to the formal
    photometric uncertainties. Values above unity indicate a
    co-variance exceeding the formal expected variance at zero lag.
    The raw light curves show clear periodicity at the 6\,hr
    (0.25\,day) spacecraft roll period. This periodicity is suppressed
    in the detrended light curves, but is still evident. The TFA light
    curves are effectively uncorrelated in time, and show no evidence
    for the 6\,hr instrumental periodicity.  }
\label{fig:autocorrelation}
\end{figure}

We find that our pre-TFA detrended light curves are comparable to those
of L16 (with a small improvement for fainter stars with $K_{p} >
14$\,mag), while the post-TFA light curves have substantially less
scatter on all timescales investigated.  Part of this is due to the
filtering by TFA of real stellar variability which is very common among
the stars in these observations, however some of this is also due to the
power of TFA at removing instrumental
systematics. Figure~\ref{fig:autocorrelation} shows the median
autocorrelation function of the light curves of bright sources with $11
< K_{P} < 12$. We find that while power at the 6\,hr roll frequency is
substantially reduced in the detrended light curve autocorrelation
function, compared to what is seen for the raw light curves, there still
remains a clear signature of these systematics even in the detrended
light curves. This signal is completely removed in the TFA light curves,
and the data are essentially uncorrelated in time.

For stars with $K_{p} > 14$ our detrended and 6.5 hour binned light
curves achieve the highest precision to date for the C0 super-stamp.
Post-processing of these light curves is still needed to remove
systematics and search for small transiting planets.  This work
represents the first published image subtraction analysis of a K2
super-stamp.  This method will provide a valuable means of analysis of
the Galactic Bulge observations carried out during K2 campaign 9.

We make the subtracted images, raw light curves, and detrended light
curves generated from the \textit{K2} C0 super-stamp publicly available
at the HAT data server: 
\url{http://k2.hatsurveys.org/archive/}.  The light curve files contain
the following information: time of observation, cadence number,
subtracted and detrended fluxes and associated errors for several
apertures, raw relative fluxes and associated errors, centroid
positions, and the accompanying PSF kernel parameters for our best
resulting light curve. The format of the light curve files is given in
Tables~\ref{table:table_TFA} and~\ref{table:tableNoTFA}, for the
detrended results (containing the data for the raw light curves therein)
and the detrended with TFA light curves, respectively.  Missing from
Table~\ref{table:tableNoTFA} (in order to save space and omit
redundancy) are the columns listing the raw and detrended rms values
associated with all 9 possible apertures. We also submit our results to
the Barbara A.~Mikulski Archive for Space Telescopes (MAST) to share
them with the scientific community.

It is our hope that there will be continued improvements of the
detrending methods and photometric analysis so that the C0 super-stamp
may be exploited to its fullest potential, including searching the data
for variable sources, which will be the subject of future work. In is
paper we have taken the first step of demonstrating that the image
subtraction method is capable of producing light curves from {\em K2}
super-stamps with a precision that is comparable to that of the best
method demonstrated to date.  We note that L16 have independently
detrended light curves and conducted a variable search on M35 and
NGC~2158 cluster members resulting in a list of 2133 variables.  For the
clusters studied here we do not expect a significantly different yield
of variables from our reduction as the light curves we have generated
are of comparable precision to those of L16.

We are currently working on extending this pipeline to other crowded
regions in the \textit{K2} field, particularly Campaign 9, which points
toward the dense Galactic Center.  We also aim to apply the pipeline to
searching for variables in globular clusters.  It is likely that image
subtraction will perform significantly better than the
neighbor-subtraction method of L16 for these particularly crowded
regions.  \textit{K2} observations have been made for a number of open
and globular clusters, including M4, M80, M45, NGC 1647, the Hyades,
M44, M67, and NGC 6717.  We aim to fully exploit these data-rich fields
using our image subtraction reduction pipeline in pursuit of new
intrinsic variables and transiting planets.

\acknowledgements
\paragraph{Acknowledgments}
MSF gratefully acknowledges the generous support from the National
Science Foundation.

The data in this paper were collected by the \textit{Kepler} mission,
which is funded by the NASA Science Mission directorate.  The data were
downloaded from the Barbara A.~Mikulski Archive for Space Telescopes
(MAST), a NASA-funded project providing astronomical data archives and
stationed at Space Telescope Science Institute (STScI).  STScI is
operated by the Association of Universities for Research in Astronomy,
Inc.


\begin{turnpage}
\begin{deluxetable*}{cccccccc}    
  \tablecaption{Detrended +  TFA Light Curve Format}
  \tablehead{\colhead{Cadence} & \colhead{Time} & \colhead{Detrended Best Ap} & \colhead{Detrended Error Best Ap} & \colhead{Raw Best Ap} & \colhead{x-coord} & \colhead{y-coord} & \colhead{TFA Detrended Best Ap} \\
    \colhead{} & \colhead{Days: BJD-2454833} & \colhead{Mag Normalized} & \colhead{Mag} & \colhead{Mag} & \colhead{Image Coords} & \colhead{Image Coords} & \colhead{Mag Normalized}}
  \startdata 
  2200 & 1940.47759875 & -0.00301 & 0.00813 & 13.96333 & 356.073 & 467.058 & -0.0004 \\
  2201 & 1940.49803044 & -0.00332 & 0.00813 & 13.96244 & 356.053 & 467.046 & 0.00025 \\
  2202 & 1940.51846192 & -0.00279 & 0.00814 & 13.96163 & 356.017 & 467.038 & 0.00045 \\
  2203 & 1940.53889351 & -0.00135 & 0.00814 & 13.96259 & 356.02 & 467.017 & 0.00227 \\
  2204 & 1940.5593252 & 0.00109 & 0.00813 & 13.96381 & 356.003 & 466.994 & 0.0018 \\
  2205 & 1940.57975669 & -0.00941 & 0.00811 & 13.95302 & 355.987 & 466.985 & -0.00864 \\
  2206 & 1940.60018828 & 0.00453 & 0.00815 & 13.96619 & 355.954 & 466.941 & 0.00154 \\
  2209 & 1940.66148304 & 0.0018 & 0.00817 & 13.9583 & 356.476 & 467.345 & -0.0015 \\
  2210 & 1940.68191473 & 0.00349 & 0.00819 & 13.96045 & 356.454 & 467.354 & -0.00051 \\
  2211 & 1940.70234632 & 0.00448 & 0.00821 & 13.96247 & 356.45 & 467.32 & 0.00134 \\
  \enddata
\label{table:table_TFA}
\end{deluxetable*}
\end{turnpage}

\tabletypesize{\scriptsize}
\begin{turnpage}
  \begin{deluxetable*}{cccccccccc}
    \tablecaption{Detrended Without TFA Light Curve Format}
    \tablehead{\colhead{Time} & \colhead{Cadence} & \colhead{Detrended Best Ap} & \colhead{Error Best Ap} & \colhead{Raw Best Ap} & \colhead{x-coord} & \colhead{y-coord} & \colhead{Arclength Param} & \colhead{Optimal PSF Kernel} \\
      \colhead{Days: BJD-2454833} & \colhead{} & \colhead{Mag Normalized} & \colhead{Mag} & \colhead{Mag} & \colhead{Image Coords} & \colhead{Image Coords} & \colhead{Image Coords} & \colhead{b-i-d--}}
    \startdata 
    1940.4776 & 2200 & -0.00301 & 0.00813 & 13.96333 & 356.073 & 467.058 & 0.6232 & b0i0d20 \\
    1940.498 & 2201 & -0.00332 & 0.00813 & 13.96244 & 356.053 & 467.046 & 0.6458 & -- \\
    1940.5185 & 2202 & -0.00279 & 0.00814 & 13.96163 & 356.017 & 467.038 & 0.6793 & -- \\
    1940.5389 & 2203 & -0.00135 & 0.00814 & 13.96259 & 356.02 & 467.017 & 0.6899 & -- \\
    1940.5593 & 2204 & 0.00109 & 0.00813 & 13.96381 & 356.003 & 466.994 & 0.7174 & -- \\
    1940.5798 & 2205 & -0.00941 & 0.00811 & 13.95302 & 355.987 & 466.985 & 0.7355 & -- \\
    1940.6002 & 2206 & 0.00453 & 0.00815 & 13.96619 & 355.954 & 466.941 & 0.7885 & -- \\
    1940.6615 & 2209 & 0.0018 & 0.00817 & 13.9583 & 356.476 & 467.345 & 0.1287 & -- \\
    1940.6819 & 2210 & 0.00349 & 0.00819 & 13.96045 & 356.454 & 467.354 & 0.1405 -- \\
    1940.7023 & 2211 & 0.00448 & 0.00821 & 13.96247 & 356.45 & 467.32 & 0.1638 & -- \\
    \enddata
    \label{table:tableNoTFA}
  \end{deluxetable*}
\end{turnpage}

\clearpage
\bibliographystyle{apj}
\bibliography{SoaresFurtado.bib}

\end{document}